\documentclass[aip,twocolumn,letterpaper]{revtex4}

\pdfoutput=1  

\usepackage{fullpage}
\usepackage{xcolor}

\usepackage{graphics}
\usepackage{epsfig}

\usepackage[margin=1.0in]{geometry}

\begin{document}
\title{ Plasma steering to avoid disruptions in ITER and tokamak power plants }
\author{Allen H Boozer}
\affiliation{Columbia University, New York, NY  10027, U.S.A.\\ ahb17@columbia.edu}

\begin{abstract}
Steering tokamak plasmas is commonly viewed as a way to avoid disruptions and runaway electrons.  Plasma steering sounds as safe as driving to work but will be shown to more closely resemble driving at high speed through a dense fog on an icy road.   The long time required to terminate an ITER discharge compared to time over which dangers can be foreseen is analogous to driving in a dense fog. The difficulty of regaining plasma control if it is lost resembles driving on an icy road.  Disruptions and runaways are associated with three  issues---a solution to one tends to complicate the solution to the other two: loss of plasma position control, excessive heat deposition, and  wall melting due to runaway electrons.  All three risks must be addressed for ITER to achieve its mission and essentially eliminated before tokamak power plants can be deployed.

\end{abstract}

\date{\today} 
\maketitle

The tokamak literature asserts that disruptions and runaways are a problem of plasma steering.  This can be found in a \emph{Physics Today} article \cite{Physics Today:2019}, which says the production of fusion energy will be enabled by the questions that ITER will answer, and in a \emph{Nuclear Fusion} article \cite{Disruption-prevention2019} reviewing progress on disruption prevention for ITER.  

The purpose of this paper is to clarify issues of plasma steering that need to be addressed for ITER to achieve its mission and for tokamak fusion-energy to be practical.  Even problems that were considered solved, such as plasma-position control when surrounded by a perfectly conducting chamber \cite{Breizman:2018,Ideal VDE,Clauser:2021}, can be more subtle than was thought.

 The steering of tokamaks to avoid disruptions is analogous to steering a car to avoid accidents.   Steering, whether  a car or a tokamak, has two fundamental problems.

The first problem for steering is foreseeing dangers.  To safely steer a car in foggy conditions, the speed of the car must be limited so it can be safely stopped within the distance at which dangerous conditions can be foreseen.  The tokamak analogue would be to limit the plasma current to a level at which it can be terminated without a disruption within the time danger can be foreseen.  

\begin{figure}
\centerline{ \includegraphics[width=3.0in]{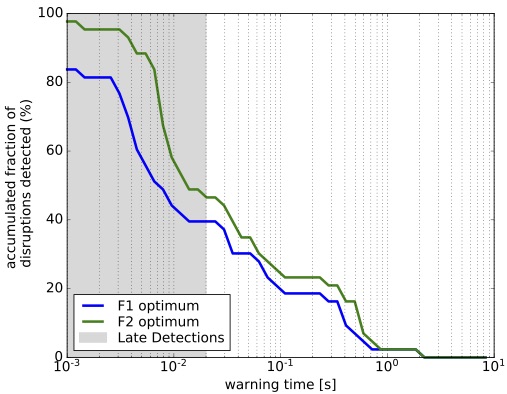} }
\caption{ Time-sliced DIII-D data was used to determine the fraction of the disruptions that were successfully predicted versus the prediction time.  F1 and F2 are two different weightings of the data.  This figure is reproduced with permission from Nucl. Fusion \textbf{59} 096016 (2019). Copyright 2019, Institute of Physics Publishing. }
\label{fig:prediction}
\end{figure}

 A review of ITER shutdown strategies \cite{ITER-shutdown2018} found that even under ideal conditions at least 60~s is required to terminate a 15 MA ITER current without a disruption.  Predictions of disruptions during the flattop period of DIII-D plasmas \cite{Granetz:2019} show a precipitous drop in reliability after milliseconds, Figure \ref{fig:prediction}, but even the short-time predictions had only modest reliability, approximately 95\%.  Many methods have been developed for predicting disruptions; a sample is  \cite{Prediction:Yokoyama,Tang:2019,Murari:2019, Kolemen:2020,Prediction:Sabbagh,Prediction:Churchill,Prediction:Zhang}.  Although results vary among the methods, the longterm predictions by all known methods have a low reliability relative to what is needed.  Even one in ten thousand pulses ending in an unmitigated disruption could have a large impact on the achievement of the ITER mission \cite{Disruption-prevention2019}.    Steering a tokamak to avoid disruptions resembles driving a car at high speed through a dense fog.  
 
The tokamak literature recognizes and discusses emergency shutdowns \cite{Disruption-prevention2019,ITER-mitigation:2016} that must be initiated orders of magnitude faster, $\sim30$~ ms.   An example, reviewed by Sertoli et al \cite{Sertoli:2014}, is a wall fragment striking the plasma.  They found that this is not a cause for disruptions on JET with an ITER-like wall but had been in other tokamak experiments.  As has long been known, wall fragments striking the plasma could become more severe in power plants because of blistering caused by alpha particles \cite{Ueda:2003}.  

A fast emergency shutdown, which means fast compared to 60~s in ITER,  requires a highly reliable strategy for instigating a benign disruption, called disruption mitigation.  But, as noted in 2019, ``\emph{With ITER construction in progress, reliable means of RE} (runaway electron) \emph{mitigation are yet to be developed}"  \cite{Breizman:2019}.  Fast shutdowns can also produce unacceptable forces on the blanket modules in ITER.  Subtleties in estimating these forces are discussed in \cite{Boozer:interaction}.

The second problem for steering is the availability and timescale of actuators, the analogs of the steering wheel and the brake pedal of a car and the 1.5~s response time of a typical driver.  For ITER the actuators are (1) the external loop voltage, (2) the externally produced axisymmetric poloidal magnetic field, (3) particle injection systems, (4) particle pumps, (5) heating and current drive systems, (6) non-axisymmetric external magnetic fields.  The major papers on plasma steering do not discuss the precise use of these six actuators, even in an unrushed shutdown \cite{ITER-shutdown2018}.  Indeed, it is unclear how to use the actuators to control what are most important for avoiding disruptions: the profile of the plasma current, the loss of position control of the plasma, and the maintenance of a sufficient plasma temperature to avoid runaways.  All of the ITER actuators except particle injection require a timescale of order seconds to be fully effective, which is too long to react to a number of envisioned situations, which require a shutdown be initiated  in of order tens of milliseconds \cite{Disruption-prevention2019,ITER-mitigation:2016}.   Even when dangers can be adequately foreseen, integration is required between the predictors and  the actuators for successful steering.  Once plasma control is lost on ITER, it is difficult to regain, much like driving on an icy road.

\begin{figure}
\centerline{ \includegraphics[width=3.0in]{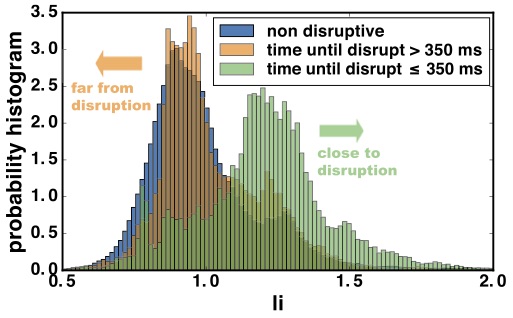} }
\caption{ Time-sliced DIII-D data was used to determine the probability that a disruption occurred within 350~ms as a function of the internal inductance $\ell_i$ during flattop periods.  Each the three histograms is normalized so that the integral under it is unity.  This figure is reproduced with permission from Nucl. Fusion \textbf{59} 096016 (2019). Copyright 2019, Institute of Physics Publishing. }
\label{fig:internal-inductance}
\end{figure}

Why does it take so long to shutdown an ITER plasma?  Magnetic fields produced outside the vacuum vessel require 0.6~s to penetrate to the plasma \cite{ITER-Vessel}, and voltage limits on the poloidal field coils typically limit large changes to times longer than several seconds  \cite{ITER-shutdown2018}.  The toroidal loop voltage on the vessel \cite{ITER-Vessel} must be less than 12~V.   At 15~MA, the poloidal magnetic flux enclosed by the ITER vacuum vessel can reach 75~V$\cdot$s, so more than 6~s would be required to remove it using the loop voltage on the vessel.  The poloidal flux removal by the resistivity of a 10~keV plasma at the magnetic axis in ITER requires $\sim 1000$~s.  Although the loop voltage on the vessel can remove the flux faster, the tendency is to produce a highly peaked current profile.  The internal inductance $\ell_i$ is a measure.  The larger $\ell_i$, the more centrally peaked the current and the greater the tendency of the plasma to disrupt, Figure \ref{fig:internal-inductance}, and the more difficult it is to keep the plasma adequately centered in the chamber \cite{ITER-shutdown2018}.  As the plasma current drops, the plasma density must be proportionately reduced to stay below the empirical Greenwald density limit \cite{Greenwald:2002}, and this requires not only particle transport out of the plasma but also particle removal from the plasma chamber.

The difficulty of benignly shutting down ITER becomes far greater during its nuclear phase than before.  Control over the power input is lost, and far more dangerous seeds for the transfer of the plasma current from thermal into relativistic electrons are present.  Even before the shutdown, steering becomes more difficult in a nuclear-powered plasma.  The current-density profile was identified in \cite{Disruption-prevention2019} as the main drive for disruptive instabilities, but which actuators ensure careful control of that profile over timescales long compared to internal flux relaxation times in a burning-plasma?  The issue may be avoided in ITER by limiting the time a plasma may be allowed to burn, but what is the solution in a power plant?

For success in a disruption-free shutdown of a burning plasma, the reduction in the plasma pressure must be consistent with adjustments to the external vertical field for the plasma to remain sufficiently centered in the machine to avoid wall contact.   Loss of centering resembles going into a skid on an icy road; regaining centering can easily become impossible.  The speed of these adjustments is strictly limited by the allowed voltages on the superconducting poloidal field coils \cite{ITER-shutdown2018}.    This is more difficult when deuterium-tritium fusion contributes 500~MW, of which 100~MW heats the plasma, with 50~MW of available external power.  The fusion power $P_{dt}$ is proportional to the plasma pressure squared within 10\% accuracy between 10 and 20~keV  \cite{Wesson:2004}.  Without a large increase in the poloidal-beta as the plasma current $I_p$ is reduced, $P_{dt}$ drops as $I_p^4$.  The effect on the plasma pressure of the precipitous drop in nuclear power as $I_p$ is decreased is magnified if the plasma switches from the high confinement H-mode to the low confinement L-mode  \cite{Disruption-prevention2019}.  

A reduction in the plasma current by a megaampere amplifies the number of energetic electrons by a factor of ten in a hydrogenic plasma \cite{Pivotal:Boozer}---even more when impurities are present \cite{McDevitt:2019,Hesslow:2019}.  In the pre-nuclear phase of ITER, the only electrons that are energetic enough to runaway  are those that were in a high-$T_e$ Maxwellian tail before the electron temperature $T_e$ was reduced sufficiently for the resistive electric field $\eta j_{||}$ to exceed the Connor-Hastie electric field \cite{Connor-Hastie}.  This is when runaway becomes possible, and at the standard ITER density requires $T_e\lesssim550~$eV.  The change from a high electron temperature $T_e\sim10~$keV to a low temperature must occur quickly, in less than the maximum collisional relaxation time of an energetic electron, the Connor-Hastie  \cite{Connor-Hastie} collision time $\tau_{ch}\approx 20~$ms.  In the nuclear phase of ITER operations, two important steady sources of energetic electrons are available: tritium decay and Compton scattering by gamma-rays from the irradiated wall, which can be amplified into dangerous relativistic-electron currents \cite{Impurity runaway:2020}.

The seriousness of disruptions and runaways is determined not only by the damage but also by the length of the shutdown required for repairs.  This is much longer after D-T operations in ITER begin.  Issues associated with ITER maintenance and repair were reviewed in 2019 by van Houtte \cite{ITER-maintenance:2019}.

Disruptions and runaways are associated with three  issues---a solution to one tends to complicate the solution to the other two: loss of plasma position control, excessive heat deposition, and  wall melting due to runaway electrons.   All three risks must be retired before tokamak power plants can be deployed.  Even the successful achievement  of the ITER mission will require not only the avoidance of disruptions in the narrow sense of a sudden loss of magnetic surfaces but also the avoidance of the production of multi-megaamperes of relativistic electrons.  Unacceptable melting \cite{Breizman:2019} can be produced by 1.9~MA of relativistic electrons striking the walls over a broad area, or 300~kA if concentrated. The risks of disruptions and runaway electrons are related but should not be conflated \cite{Boozer:interaction}.  In particular, the avoidance of magnetic-surface breakup can exacerbate the risk of runaway electrons.

Fusion has the potential of making a major contribution to stopping the increase in atmospheric carbon dioxide  \cite{CO2-Stell,NASEM:2021}.  For this, minimization of time and risk for a demonstration of fusion power is of great importance.  The United States National Academy of Sciences, Engineering, and Medicine stated \cite{NASEM:2021}: ``\emph{the Department of Energy and the private sector should produce net electricity in a fusion pilot plant in the United States in the 2035-2040 timeframe.}"  The cost of each year's delay in developing a solution, approximately a trillion dollars  \cite{CO2-Stell},  far exceeds the credible cost of a minimal time and risk fusion program.  

The cost of deploying a sufficient number of fusion reactors to have a significant effect on carbon dioxide production is order a thousand times greater than constructing a demonstration fusion power plant.  Nevertheless, having one working fusion power plant is important in itself to world security.  The precise cost of fusion energy is only relevant during the deployment phase in comparison with other solutions---and each of the alternatives for a complete energy system has major disadvantages in comparison to fusion  \cite{CO2-Stell}.  The cost of electricity and the minimum unit size are only two considerations.  Others can be more important: intermittency, site specificity, waste handling, and the potential for nuclear proliferation.

Making the risks of disruptions and runaways acceptable in ITER is difficult but far easier than in DEMO, a machine that can demonstrate fusion power \cite{DEMO-challenges}.  The basic problem is the structures surrounding the plasma are more delicate in a power plant than they are in ITER.  In addition, the diagnostics, which are needed for steering, become much more limited \cite{Diagnostics:2019}.

Magnetic fusion systems can be designed to be robust against disruptions and runaways by making them non-axisymmetric \cite{CO2-Stell}.   As stated by the U.S. National Academy \cite{NASEM:2021}, an assessment of fast paths to fusion energy requires a multi-year design study of potential fusion power plants.

Disruption and runaway issues are far more challenging in a tokamak power plant than during D-T operations of ITER and far more challenging in D-T operations of ITER than in non-D-T operations.  A demonstration that disruption and runaway issues can be adequately addressed for practical tokamak fusion power will have to wait approximately thirty years until this can be demonstrated by power plants having operated an adequate period of time.  A negative conclusion on practicality could come sooner: after fifteen years when D-T operations start on ITER or after five years when ITER starts plasma operations.  In a 2021 paper, Nicholas Eidietis recognizes the challenges that disruptions pose to tokamak power plants but remains optimistic that these challenges can be met \cite{Eidietis:2021}.

Careful thought is required to determine how timescales should be integrated within an overall fusion program designed to minimize risk and time in demonstrating fusion power at the level required for informed decisions on its deployment.  Each year's delay in deploying carbon-free energy systems not only costs of order a trillion dollars \cite{CO2-Stell} but also affects security worldwide.

\vspace{0.2in}

\section*{Acknowledgements}

This material is based upon work supported by the U.S. Department of Energy, Office of Science, Office of Fusion Energy Sciences under Award Numbers DE-FG02-03ER54696, DE-SC0018424, and DE-SC0019479.   





\end{document}